\documentstyle[12pt,fleqn]{article}
\def\beq{\begin{equation}}
\def\eeq{\end{equation}}
\def\Eq{Eq.~(\ref}
\def\1{\lambda}
\def\6{\langle}
\def\9{\rangle}
\def\bB{\mbox{\bf B}}
\def\bC{\mbox{\bf C}}
\def\bP{\mbox{\bf P}}
\def\bZ{\mbox{\bf Z}}

\begin{document}

\vspace*{10mm}
\begin{center}
{\Large{\bf All the Bell Inequalities}}\\[2cm]
Asher Peres$^*$ \\[7mm]
{\sl Department of Physics, Technion---Israel Institute of
Technology, 32\,000 Haifa, Israel}\end{center}\vfill

\noindent{\bf Abstract}\bigskip

Bell inequalities are derived for any number of observers, any number of
alternative setups for each one of them, and any number of distinct
outcomes for each experiment. It is shown that if a physical system
consists of several distant subsystems, and if the results of tests
performed on the latter are determined by local variables with objective
values, then the joint probabilities for triggering any given set of
distant detectors are convex combinations of a finite number of Boolean
arrays, whose components are either 0 or 1 according to a simple rule.
This convexity property is both necessary and sufficient for the
existence of local objective variables. It leads to a simple
graphical method which produces a large number of generalized
Clauser-Horne inequalities corresponding to the faces of a convex
polytope. It is plausible that quantum systems whose density matrix has
a positive partial transposition satisfy all these inequalities, and
therefore are compatible with local objective variables, even if their
quantum properties are essentially nonlocal.\vfill

\noindent $^*$\,Electronic address: peres@photon.technion.ac.il\vfill

\noindent Published in {\it Foundations of Physics\/} {\bf29}, 589--614
(1999).

\newpage\noindent{\bf1. \ PATTERNS OF LOCAL OBJECTIVE VARIABLES}\medskip

\noindent A local realistic theory purports that the results of tests
(or ``measurements'') performed on distant physical systems by
independent observers are determined by local variables with objective
values. It was shown long ago by Bell~[1] that this assumption led to an
upper bound on the correlations of the results of these tests, and that
this bound was violated by the predictions of quantum theory. Early
versions of Bell's inequality~[1,~2] involved only two observers, each
one having a choice of two (mutually incompatible) experiments. The
various outcomes of each experiment were lumped into two sets,
arbitrarily called +1 and $-1$. Generalizations involving more than two
observers, or more than two alternative experiments for each observer,
or more than two distinct outcomes for each experiment, were considered
by many authors~[3]. Another type of inequality was derived by Clauser
and Horne~[4], who supplied only one detector to each observer, so that
not all events could be detected. This simpler inequality has the
advantage that it is not affected by our lack of knowledge of the total
number of quantum systems, which in general is not observable. Still,
detector inefficiencies reduce the chances of observing an actual
violation of the CH inequality.

In the present article, it will be shown that all these inequalities can
be derived from a universal convexity rule, which is not only a
necessary condition, but also a sufficient one for the existence of
local objective variables. To avoid a possible misunderstanding: this
article is not an attempt to attribute physical reality to the so-called
``hidden'' variables, nor to speculate on their dynamical properties. It
is only an intellectual exercise extending the work of Bell and others
in the most general way. The only assumption is that, for a given set
of physical situations, there exist local rules that determine
coincidences between detections of distant macroscopic events. This
assumption is sufficient to derive constraints on the probabilities of
these coincidences, in the form of inequalities. 

Probabilities can in principle be measured experimentally with arbitrary
accuracy. A violation of any one of the above constraints is sufficient
for proving the non\-existence of local hidden variables (that is, of
hypothetical local rules as postulated above). Conversely, if all the
constraints are satisfied for a particular physical preparation and a
particular set of detectors, then such rules can be formulated. The sole
subject of this article is to define what is meant by {\it all\/} the
constraints (i.e., all the Bell inequalities.)

The first step of this work is to develop suitable algebraic and
graphical notations for representing all we shall need to know about
local objective variables. Then, in Sect.~2, it is shown that every Bell
inequality can be written as $\sum F^KP_K\ge0$, where the $P_K$ are
probabilities of detector coincidences (which can be measured
experimentally), and the coefficients $F^K$ are a set of integers. The
problem is to find these coefficients explicitly. A method generating
sets of $F^K$ for any number of observers, any number of alternative
setups for each one of them, and any number of distinct outcomes for
each experimental setup, is given in Sect.~3. Sequential tests and
quantum distillation are discussed in Sect.~4, and the main new results
in this article are summarized in Sect.~5. Finally, an appendix presents
an algebraic (rather than graphical) method for constructing the vectors
$F^K$.

Consider first the case of two observers, conventionally called Alice
and Bob (the presence of additional observers will be discussed whenever
appropriate). Alice has a choice of several tests, which may be mutually
incompatible. The fact that some tests are mutually incompatible is a
consequence of the quantum formalism (Bohr's principle of
complementarity). However, quantum theory is not involved at all in the
present discussion, which is strictly phenomenological. All we have to
know is that Alice's first type of test, if performed, yields one of
several distinct outcomes, labelled $a$, $b$, $c$, etc. Her second type
of test may likewise yield outcomes labelled $m$, $n$, $s$, etc., her
third type of test has outcomes $u$, $v$, $w$, and so on. These symbols
are completely arbitrary, for example, they are labels marked on Alice's
detectors. Note that Alice is free to choose the test that she performs.
It is the result of the test that is in general not known to her in
advance.

Likewise, Bob has a free choice of performing a test whose possible
outcomes are $\alpha$, $\beta$, $\gamma$, etc., or another test
(perhaps incompatible with the first one) with outcomes labelled $\mu$,
$\nu$, $\sigma$, and so on. Additional observers, if any, would also
have a free choice between various, possibly incompatible tests. It is
essential that each observer can choose his/her test independently of
what the other distant observers do (or have done, or will do).

Local hidden variable theories assume that each preparation of the
physical system is characterized by a parameter $\1$, which acts as a
kind of ``set of instructions''~[5] that determines the outcome of any
test that can be performed on each one of the subsystems. For example,
as illustrated in Fig.~1, if Alice has a choice of three tests, a
particular value of $\1$ would determine that their outcomes are $b$,
$s$, or $u$ (whenever Alice elects to perform the corresponding test),
and likewise the outcomes of Bob's two possible tests are $\beta$ or
$\rho$. Even if present technology does not allow us to control the
value of $\1$, we may speculate that such ``hidden variables'' $\1$ do
exist, and the statistical results of our coarse experiments correspond
to an average over a suitable probability distribution for $\1$. Note
that if a test fails to provide a detection event because of detector
inefficiency (or lack of detector in the appropriate direction), this
also has to be considered as an event in the statistics. Therefore each
sector in Fig.~1 should include one row and one column labelled~0 (or
likewise). These merely correspond to hidden variables $\1$ whose ``set
of instructions'' is: no detection.

It is crucial for the following argument to have a clear system of
notations. We shall associate to each coincidence of outcomes, such as
$b\beta$, or $u\sigma$, etc., a single symbol $K$, which stands for a
pair of indices (one index for each observer). The values of $K$ thus
label all possible coincidences between the detectors that are used by
the two observers, in all their alternative experimental setups. If
Alice's $i$-th test has $A_i$ distinct outcomes, and Bob's $j$-th test
has $B_j$ distinct outcomes, there are

\beq N_K=\sum_i A_i\,\sum_j B_j \label{NK} \eeq
distinct values of $K$ (i.e., types of coincidences between the
observers' detectors).

Note that to each value of the hidden variable $\1$ corresponds a well
defined set of values of $K$. For example, Fig.~1 illustrates a
particular $\1$ which is associated with the coincidences $b\beta,\
b\rho,\ s\beta,\ s\rho,\ u\beta$, and $u\rho$. We could therefore denote
this $\1$ more explicitly by a symbol such as $[bsu;\beta\rho]$. In
general, each $\1$ refers to a definite pattern of points in the figure.
This means that effectively there are no more than

\beq N_\1=\prod_i A_i\,\prod_j B_j \eeq
different values of $\1$, covering all the combinations of experimental
setups used by Alice and Bob. We don't have to speculate about the
physical nature and dynamical properties of the hidden variables. For
our present purpose, we may simply lump together all the values of the
hidden variables that determine the same pattern of outcomes, and denote
them collectively by a single label $\1$.  All these notions are readily
generalized to the case of three or more observers. Each new observer
should have a private set of labels for the outcomes of his/her tests.
The collective index $K$ (which indicates detector coincidences)
comprises one individual index for each observer. The case of three
observers is illustrated in Fig.~2, where a typical value of $K$ would
be $\gamma m{\sf R}$.

We now define a Boolean matrix $B_K^\1$ as follows: $B_K^\1=1$ if the
hidden variable $\1$ leads to a coincidence of type $K$ (whenever the
various observers choose to perform the kind of test that may generate
that type of event), and $B_K^\1=0$ if a coincidence of type $K$ never
results from that particular $\1$. Explicitly, when the collective
indices are written in full detail, we have, for example,

\beq B^{asu;\beta\rho}_{m\mu}=(\delta_m^a+\delta_m^s+\delta_m^u)
 \,(\delta_\mu^\beta+\delta_\mu^\rho). \label{Bfactor}\eeq
Note that this is a product of two factors, each factor related to one
of the observers. The importance of this property will become apparent
later (see Appendix).

In general, if Alice has a choice of $N_A$ tests, and Bob a choice of
$N_B$ tests, and so on, the total number of different tests (i.e., of
sectors in a diagram like Fig.~1) is

\beq N_T=N_A\,N_B\,\ldots\ , \eeq
and we have, for each value of $\1$,

\beq \sum_K B_K^\1=N_T. \label{NT}\eeq
The $N_\1$ different values of $\1$ are nothing more than labels which
distinguish various combinations of events (i.e., various patterns of
points in Fig.~1). It is convenient to define ``vectors'' $\bB^\1$ whose
components are $B_K^\1$ (where $\1$ is a ``name'' identifying each
vector, and $K$ is an index referring to each one of its $N_K$
components). These vectors are completely defined by enumerating the
tests available to each observer and the outcomes of each one of these
tests, as explicitly shown by \Eq{Bfactor}). The uniqueness of this set
of vectors is crucial for proving the convexity property expressed in
\Eq{cnvx}) below.

An important property of the vectors $\bB^\1$ is that they are not
linearly independent. The linear dependence is due to the existence of
equivalent superposition of patterns. For example, it follows from
\Eq{Bfactor}) that

\beq B^{asu;\beta\rho}_{m\mu}+B^{asu;\alpha\sigma}_{m\mu}=
 B^{asu;\beta\sigma}_{m\mu}+B^{asu;\alpha\rho}_{m\mu}.\eeq
Many more such relationships appear when a larger number of indices are
permuted. In spite of these numerous linear dependence relations, none
of the vectors $\bB^\1$ can be expressed as a {\it convex\/} combination
of the others (that is, a linear combination with positive
coefficients). This is easily seen from the fact that each one of these
vectors has exactly $N_T$ components equal to~1, and all the others
are~0. Therefore, each vector $\bB^\1$ is an extreme ray of a convex
cone.

Finally, let us note that hidden variables as described above have a
deterministic nature, because each $\1$ completely specifies the
outcomes of all experiments that can be performed. It is also possible
to imagine {\it stochastic\/} hidden variables, which only attribute
definite probabilities to these outcomes~[6]. In that case, the values
of $B^\1_K$ are arbitrary positive numbers, summing up to~1 in each
sector. This obviously is equivalent to subdividing the hidden variables
$\1$ into other, more detailed, deterministic hidden variables. No new
Bell inequality can be obtained in this way.\bigskip

\noindent{\bf2. \ FARKAS'S LEMMA}\medskip

\noindent Different $\1$ are just like different pages in a catalog~[7]
which specifies the outcomes of all experiments that may follow a given
preparation. Let $w_\1$ be the probability of occurrence of a particular
$\1$. Let each observer choose a specific test. In this way, one of the
sectors in Fig.~1 is selected, and each value of $\1$ determines the
resulting coincidence, $K$. The probability of that outcome is

\beq P_K=\sum_\1 w_\1\,B_K^\1.\label{cnvx} \eeq
The left hand side of this equation is an experimentally accessible
quantity; it may also be computed by quantum theory, or any other
supposedly valid theory. Note that $\sum_K P_K=N_T$, since there are
$N_T$ possible combinations of tests.

In geometrical terms, Eq.~(\theequation) states that the vector \bP\
(whose components are the probabilities $P_K$) is a convex combination
of the known Boolean vectors $\bB^\1$~[8,~9]. This obviously is a
necessary condition for the existence of local objective variables $\1$
as defined above. This is also a sufficient condition: if the vector
\bP\ lies within the convex cone of the given vectors $\bB^\1$, then
it is possible to expand $P_K$ as in Eq.~(\theequation), with
non-negative coefficients $w_\1$. It is even possible, in general, to do
that in an infinity of ways, if $N_\1>N_K$. 

Even though this problem is in principle straightforward, it is
notoriously difficult to determine whether or not a specific vector is
included in a given convex cone. A necessary condition can be derived
from Eq.~(\theequation) as follows.  Let a vector with real components
$F^K$ have the property that, for all $\1$,

\beq M\le\sum_K B^\1_K\,F^K\le N, \label{Farkas}\eeq
where $M$ and $N$ are fixed numbers. Then, obviously, since the
$w_\1$ are non-negative and sum up to 1, we also have

\beq M\le \sum_K P_K\,F^K\le N. \eeq
In the special case $M=0$, it can be proved (this is Farkas's
Lemma~[10]) that if the above relationship holds for {\it every\/}
vector $F^K$ that satisfies

\beq \sum_K B^\1_K\,F^K\ge0\qquad \forall\1, \label{Farkas0}\eeq
then the inequality

\beq \sum_K P_K\,F^K\ge0\label{PF}\eeq
is a {\it sufficient\/} condition for \bP\ to be in the convex cone
spanned by the vectors $\bB^\1$, in accordance with \Eq{cnvx}).
Obviously, the Farkas vectors with components $F^K$ also form a convex
cone. It is in principle possible to find all its extreme rays by
algebraic methods, as shown in refs.~[8,~9] and here in the Appendix.
Unfortunately, that algebra becomes quite unwieldy when $N_\1$ is large.
It requires an amount of computation which increases exponentially with
$N_\1$, and therefore it has little practical value. The main purpose of
this article is to derive a graphical method which generates a large
number of Farkas vectors (admittedly, only the simplest of these graphs
are explicitly displayed here, and I make no claim of completeness).

The importance of Farkas vectors in the present context stems from the
fact that the relationship (\ref{PF}) is not trivial, if some of the
components $F^K$ are negative. The probabilities $P_K$ can, in
principle, be measured experimentally with arbitrary accuracy. Thus, if
it turns out that \Eq{PF}) is not satisfied by the experimental data, it
follows that the assumption of existence of local objective variables is
incompatible with physical reality. Bell's inequality and all similar
relations~[1--4] are special cases of \Eq{PF}).

I shall now present a simple graphical method which produces a large
number of Farkas vectors with $M=0$ (the value of $N$ is irrelevant,
since it can always be adjusted by multiplying all the components $F^K$
by a suitable positive factor). That set of Farkas vectors may be
redundant, but it is finite.  These explicit results are far
preferable to the algebraic algorithm described in the Appendix. The
latter is hopelessly inefficient because it is too general: it makes no
use of the peculiar structure of $B^\1_K$, given by \Eq{Bfactor}).

Let us introduce more explicit notations and replace the composite index
$K$ by its individual components. Moreover, let us use the term
``vector'' with the meaning ``set of components'' if no confusion may
arise. For example, if there are two observers, the Farkas vector $F^K$
may be written as $F^{m\mu}$. Likewise, for each hidden variable label
$\1$, the vector $B^\1_K$ can be written as $B^\1_{m\mu}$. Such a
notation is sometimes useful, as in Eqs.~(\ref{minus}) and (\ref{plus})
below. Moreover, it leads to a graphical method with considerable
intuitive appeal: each one of these vectors is represented by a {\it
rectangular array\/}, where rows are labelled by Alice's Latin indices,
and columns by Bob's Greek indices. (The generalization to $N$ observers
is obvious: each vector becomes an $N$-dimensional array.)

The stucture of the $B^\1_K$ arrays is clear. For each $\1$, the array
looks like the diagram in Fig.~1, with +1 at each node of the pattern,
and zeros everywhere else. With these notations for $B^\1_{m\mu}$, it is
easy to construct some Farkas vectors. They can be represented by
rectangular patterns as in Fig.~3 and~4, with black cells and gray cells
corresponding to components $F^K$ whose values are $-1$ and +1,
respectively (if color printing were readily available, I would have
used red and green cells with an obvious meaning, as in traffic lights).
Any empty cell stands for a component $F^K=0$. Other values, if needed,
would have to be explicitly written at the appropriate location in each
pattern.

The construction of nontrivial Farkas vectors will be discussed in the
next section. As a preliminary step, some ``trivial'' ones are mentioned
below, because they are necessary for future use.

Farkas vectors like those in Fig.~3 are denoted as $Z^K$. They have
components

\beq Z^{a\mu}=Z^{b\mu}=\cdots=-1,\label{minus}\eeq
filling a complete line in one sector, and

\beq Z^{r\mu}=Z^{s\mu}=\cdots=1,\label{plus}\eeq
filling the continuation of the same line, in one other sector.
Obviously, any overlap of a $\1$-pattern with such a Farkas vector will
include one negative cell and one positive cell, so that

\beq \sum_K B_K^\1\,Z^K\equiv0. \label{ZK} \eeq
We thus expect the experimental data to fulfill

\beq \sum_K P_K\,Z^K=0. \label{PF0}\eeq

Indeed, this equation is always satisfied. This can be seen as follows.
The left hand side of (\theequation) consists of sums of coincidence
probabilities

\beq p_n=\sum_\gamma p_{n\gamma}\qquad\qquad\mbox{\rm and}\qquad\qquad 
 p_\mu=\sum_c p_{c\mu}, \label{single} \eeq
where the summation indices $\gamma$ and $c$ take all the values in {\it
any one\/} of the sectors. One sum appears with negative signs, as in
\Eq{minus}), and one with positive signs, as in~(\ref{plus}). The sums
$p_n$ and $p_\mu$ are single detector probabilities which can be
measured by Alice and Bob, far away from each other. It is known fact
that probabilities measured by one observer do not depend on the {\it
choice\/} of experiments performed by other, distant observers (if it
were such a dependence, it could be used for instantaneous signaling at
arbitrary distances, contrary to relativity theory and to common
experience).  Therefore the sums in Eq.~(\theequation) have to be the
same, irrespective of the sector that was chosen to evaluate them, and
\Eq{PF0}) is always satisfied by the experimental data.

It follows that if $F^K$ is a Farkas vector, then $G^K=F^K+aZ^K$ (for
any real $a$) also is a Farkas vector. These two vectors are physically
equivalent. If $\sum P_KF^K\ge0$ is experimentally satisfied (or
violated), so is $\sum P_KG^K\ge0$, and vice versa. In the rest of this
article, we shall therefore consider only equivalence classes of Farkas
vectors, defined modulo arbitrary linear combinations of the various
$Z^K$.

The existence of null vectors complicates the definition of extreme rays
in the set of Farkas vectors. Since both \bZ\ and $-\bZ$ are Farkas
vectors, the notion of a convex combination becomes ambiguous. Together
with each ray, we have to consider its entire equivalence class: an
extreme ray can then be defined as one that cannot be obtained by convex
combinations of rays taken from {\it other\/} equivalence classes. A
different definition of extremality will be given in the Appendix, where
we shall work in the subspace spanned by the Boolean vectors $\bB^\1$. A
Euclidean metric will be introduced in $K$-space, and we shall construct
Farkas vectors that are orthogonal to all the \bZ\ vectors.\bigskip

\noindent{\bf3. \ CONSTRUCTION OF FARKAS VECTORS}\medskip

\noindent Farkas vectors can be classified as those having no negative
component (these vectors are trivial and useless), or negative
components in a single sector, as those in Fig.~3 or~4, and so on. It is
easily seen that it is not necessary to consider the possibility of
negative components in more than one sector. Indeed, if such a situation
happens, it is always possible to concentrate all the negative
components into a single sector by adding appropriate null Farkas
vectors $Z^K$. Therefore we shall assume that only one sector has
negative components. Moreover, we can assume that it has only negative
or null components, because any positive component in that sector,
combined with positive components in the other ones, would never lead to
a contradiction with \Eq{PF}), since each $\1$-pattern touches each
sector only once. It will be shown in the Appendix that Farkas vectors
can be chosen in such a way that all their components are integers.

Let us first consider the simplest case: two observers, two alternative
setups for each one of them (so that each array of $B^\1_{m\mu}$ has
four sectors), and two outcomes, or sets of outcomes, for each test
(i.e., each sector has four cells). The various outcomes will now be
labelled by symbols such as $m$ and $m'$. The latter means ``not $m$''
and may include the null outcome (no detection).

There are 64 Farkas vectors like the one drawn in Fig.~4, since each
one of the 16 squares can be chosen as the black one ($-1$), and then
any one of the four squares in the opposite sector can be made gray
(+1). The two other gray squares in the remaining sectors then have
fixed locations: each one is in the same row or column as the black
square, and none is in the same row nor column as the other gray
squares. This structure guarantees that if any Boolean pattern (as in
Fig.~1) overlaps with the negative component, it must also overlap with
either one or two positive components (so that the sum is 0 or 1).
Moreover, no Boolean pattern overlaps with two positive components
without a compensating $-1$. For the vector represented in Fig.~4, we
thus obtain from \Eq{PF})

\beq 0\le p_{a\mu'}+p_{n'\beta}+p_{n\mu}-p_{a\beta}\le1.
 \label{Fineq}\eeq
This result is equivalent to the Clauser-Horne inequality~[4]. Indeed,
by adding and subtracting $(p_{n\beta}+p_{a\mu})$, we obtain, by
virtue of \Eq{single}),

\beq 0\le p_a-p_{a\mu}+p_\beta-p_{n\beta}+p_{n\mu}-p_{a\beta}\le1,
 \label{CH} \eeq
which is the CH inequality~[4]. The more familiar CHSH inequality~[2]
can easily be derived from it, but the converse is not true, unless all
detectors have perfect efficiency (that is, no events are undetected).
Note that primed indices, which include the ``no detection'' outcome,
are absent from Eq.~(\theequation).

It is only the lower limit in the above inequalities that is relevant to
Farkas's lemma. This is an important feature of the CH inequality,
because detectors do not measure probabilities---there just count
events. To obtain probabilites, we have to divide these counts by the
total number of quantum systems that are tested, and that number is
never known (it can only be estimated theoretically). It is therefore
important that the lack of knowledge of that number does not affect the
lower limit of the CH inequality.

The Farkas vector depicted in Fig.~4 is not the only one for the
$2\times2$ case. Another one is shown in Fig.~5. However, the latter is
easily seen to be equivalent to a trivial Farkas vector without negative
elements, and it is therefore irrelevant.\bigskip

\noindent{\bf3.1. \ More than two detectors}\\[3mm]
If Alice or Bob have more than two distinct outcomes in an experimental
setup, the number of $\1$-patterns increases, and so does the number of
Farkas vectors. An obvious consistency requirement is that, if all Bell
inequalities are satisfied for a given set of detectors, and if the
outputs of several of these detectors are merged, as if they were a
single detector, then the new Bell inequalities for the simplified data
analysis are still satisfied. Conversely, if some Bell inequality is
violated in a setup that includes a coarse grained detector, and if the
latter is subdivided into several detectors with higher resolution,
there will still be, in the new setup, at least one Bell inequality
which is violated.

Both consistency requirements are manifestly fulfilled by Farkas vectors
of the type depicted in Fig.~6. In each experimental setup, the various
detectors may be grouped into two sets in all possible ways. If Alice's
two tests have $A_1$ and $A_2$ distinct outcomes, and if Bob's two tests
have $B_1$ and $B_2$ outcomes, there are

\beq N_F=(2^{A_1}-2)\,(2^{A_2}-2)\,(2^{B_1}-2)\,(2^{B_2}-2) \eeq
different Farkas vectors that can be constructed by this method. Each
one generates a generalized CH inequality.

Is this set of Farkas vectors complete? It is plausible, but I have no
formal proof, that any other Farkas vector is equal to a convex
combination of those constructed by the above method, plus any trivial
vectors having only positive components, plus any linear combination of
the irrelevant vectors $Z^K$. In other words, all the extreme points of
the convex set of Farkas vectors are included among those constructed as
in Fig.~6. Note in particular that it is useless to introduce complete
lines of negative elements: these can be eliminated by adding suitably
chosen null vectors $Z^K$, as illustrated in Fig.~7.\bigskip

\noindent{\bf3.2. \ More than two alternative tests}\\[3mm]
\noindent If Alice or Bob have more than two different experimental
setups to choose from, each $\1$-pattern has more than four sectors, and
therefore some Farkas vectors may have nonvanishing components in more
than four sectors. For instance, if Alice and Bob can choose from 3
setups, there are 9 sectors in each diagram, as seen in Fig.~8. Any
$\1$-pattern $B^\1_{m\mu}$ indicates 9 coincidences, located at the
intersections of any 3 rows and 3 columns.

As an example, the diagram on the left hand side of Fig.~8 represents a
Farkas vector which produces a ``chained'' CH inequality~[11]. This is
indeed a Farkas vector, because if a $\1$-pattern overlaps with the
black square ($-1$) , it has to overlap with at least one gray square
(+1), so that the sum is never negative (this is easily seen by direct
inspection of the figure). However, it is also easily seen that this
Farkas vector can be decomposed into a sum of two simpler Farkas
vectors, each one involving only four sectors. This decomposition is
like the algebraic identity

\beq ab+bc+cd+de+ef-fa\equiv (ab+bc+cd-da)+(ad+de+ef-fa), \eeq
and it can be extended to longer chains, in the same way.

Here, we may be tempted to speculate that any Farkas vector spread over
more than four sectors can be decomposed into a sum of Farkas vectors
involving four sectors only. Namely, if a Bell inequality involving the
results of more than four incompatible experiments is violated, then
there is another Bell inequality, involving only four of these
experiments, which is also violated. Unfortunately, there are
counter\-examples, namely sets of $P_K$ that invalidate this
conjecture~[12,~13]. It happens that these counter\-examples are
incompatible with quantum mechanical probabilities for spin-$1\over2$
particles~[14], but counter\-examples also exist for genuine quantum
probabilities in higher dimensional Hilbert spaces~[15]. Anyway, the
present study is not about the values of quantum mechanical
probabilities, but about general structural relations among observable
probabilities deriving from local objective variables.

It would be useful to have an efficient algorithm for constructing all
the Farkas vectors for any number of sectors. We know that this can be
done in such a way that one of the sectors has no positive elements, and
all the others have no negative elements. I have not yet been able to
find such an algorithm, and at this time I can only say that the
construction of Farkas vectors is a simple combinatorial problem,
amenable to a computer search if the number of sectors is not too large,
and if the values of the components $F^K$ are restricted to small
integers.  Note that any optimization~[16] of the experiments testing
the CH or CHSH inequalities is equivalent to having an infinite number
of sectors, but considering only Farkas vector that lie in four
sectors.\bigskip

\noindent{\bf3.3. \ More than two observers}\\[3mm]
\noindent A three-dimensional pattern of hidden variables, for
measurements performed on three correlated subsystems, was illustrated
in Fig.~2. In such a case, Farkas vectors are also represented by
three-dimensional arrays, as for example in Fig.~9. Each cubic element
corresponds to a coincidence of three detectors, and each cubic sector,
separated from other sectors by thick lines, represents a complete
experimental setup. The simplest null vectors \bZ\ correspond to a {\it
line\/} of cubes (not to a {\it slab\/} of cubes). Each component $Z^K$
refers to a specified coincidence of detections by two of the observers,
irrespective of the result found by the third observer.  By the same
argument as before, it is possible to transfer all the negative elements
of a Farkas vector into one of the 3-dimensional sectors, so that the
other sectors have only positive or null elements.

The obvious generalization of the Farkas diagram for the CH inequality
(Fig.~4) is the cubic array shown in Fig.~9. Note that only one slab of
that array is used: the negative element (shown as a black cube) is
always compensated by at least one of the positive elements (the gray
cubes) in the same slab. This means that the third observer does not
need several incompatible experimental setups. (Still, on Fig.~9, eight
sectors have been drawn, indicating that the third observer has a choice
between two alternative setups, but actually his second setup is never
used.) The data analysis for the Farkas vector of Fig.~9 considers only
those cases where the third observer gets a particular outcome (namely,
the outcome corresponding to the first slab in the cubic array). If he
gets any other outcome, the two other observers discard their parts of
the composite quantum system (without testing them), and the three
observers proceed with the next trio of entangled particles.

As a simple example, borrowed from quantum theory, consider the
3-particle state

\beq \psi_{123}=(x_1\,x_2\,x_3+y_1\,y_2\,y_3)/\sqrt{2},\label{qm} \eeq
where $x_j$ and $y_j$ are two orthogonal states of the $j$-th particle.
A Hadamard transform,

\beq x=(u+v)/\sqrt{2},\qquad\qquad y=(u-v)/\sqrt{2},\eeq
gives

\beq \psi_{123}=(u_1\,u_2\,u_3+u_1\,v_2\,v_3+
 v_1\,u_2\,v_3+v_1\,v_2\,u_3)/2.\eeq
Let the first observer test whether the state of his particle is $u_1$.
If the answer is affirmative, the resulting (renormalized) state of the
two other particles is 

\beq \psi'_{23}=(u_2\,u_3+v_2\,v_3)/\sqrt{2}.\label{psi23}\eeq
This is a maximally entangled state, violating the CH inequality.
Similar results are obtained if any one of the observers elects to test
his particle for the state $u_j$, or for the orthogonal state $v_j$.
When the test succeeds, the particles of two other observers are
maximally entangled. We see here the advantage of the CH
inequality~(\ref{Fineq}).  It is only the left hand side of that
inequality ($0\leq\ldots\ $) that is needed for Farkas's lemma, and that
side is not affected if all the probabilities that appear in it are
renormalized by the same factor (namely, they are divided by the
probability that the preliminary test was successful).

All these considerations are readily generalized to $N$ observers
($N>3$). Only two of the observers need to have access to alternative
setups. All the other ones just have to select one particular outcome of
a fixed test. How to best choose that test, for example how to orient
the detection apparatus of each observer, depends on the theoretical
model that is under consideration (in the preceding example, if an
observer would test for states $x$ or $y$, instead of $u$ or $v$, the
resulting state of the two other particles would not be entangled).
However, finding optimal tests is {\it not\/} the subject of the present
article. Here I assume that the various tests have been specified, and
the only problem is how to analyze the resulting data.

The above method may be contrasted with those of other authors~[17--19]
who sought to generalize the CHSH inequality~[2]. In all these
generalizations, an exponentially large number of tests ($\sim2^N$) are
performed. Violations of the generalized CHSH inequalities can also be
exponentially large, but their detection is subject to an exponentially
large noise~[20]. The method proposed here is clearly preferable.
However, as in earlier cases, I must acknowledge that I have no formal
proof that the above construction, together with all its variants
obtainable by permuting the various detectors, includes all the extreme
points of the convex set of Farkas vectors. Namely, it still remains to
be formally proved that if all the CH inequalities constructed by the
above method are satisfied for a given composite system and a given set
of detectors, then there is no way of combining the data of the same
detectors so that a Bell inequality is violated.\bigskip  \clearpage

\noindent{\bf4. \ SEQUENTIAL TESTS AND QUANTUM DISTILLATION}\medskip

\noindent Quantum mechanics was first mentioned above in \Eq{qm}).
Its relationship to classical mechanics is fairly well understood for
systems with continuous variables. However, the Bell inequalities
involve only classical logic for Boolean variables, and the relation of
these inequalities with quantum mechanics ought to be clarified.

There is one property that is well known: if a quantum state is
factorable, namely $\rho=\rho'\otimes\rho''$ where $\rho'$ and $\rho''$
refer to Alice's and Bob's particles, respectively, then the CH
inequality~(\ref{CH}) is always satisfied. This is easily proved as
follows. First, we note that in such a case the coincidence
probabilities also factor, e.g., $p_{a\mu}=p_ap_\mu$. Let
$j=1,\ldots,N$ be a label indicating the $j$-th experimental run.
Define $a_j=1$ or 0 if the detector labelled $a$ is excited or is not
excited, respectively, in the $j$-th run; and let there be similar
notations for the other detectors. Now consider four numbers, $a_j,\
n_j,\ \beta_j,\ \mu_j$, each one 0 or 1. It is easily verified, by
direct inspection, that

\beq 0\leq
 a_j\,(1-\mu_j)+\beta_j\,(1-n_j)+n_j\,\mu_j-a_j\,\beta_j\leq1. \eeq
Averaging this expression over all runs, the inequality~(\ref{CH})
readily follows. More generally, the CH inequality is satisfied by any
convex combination of factorable states,

\beq \rho=\sum_i c_i\,\rho'_i\otimes\rho''_i, \label{separ}\eeq
with $c_i>0$ and $\sum_i c_i=1$. Such a state is called separable. On
the other hand, there are demonstrably inseparable states that do not
violate any Bell inequality~[21,~22].  Thus, although Bell inequalities
guarantee the possibility of introducing local objective variables (for
the results of the tests under consideration), they do not guarantee
quantum separability.

A simple necessary condition for separability can be obtained as
follows~[22]. Given the density matrix $\rho$ in \Eq{separ}), consider
the partly transposed matrix

\beq \sigma=\sum_i c_i\,\rho'_i\otimes\tilde{\rho}''_i, \eeq
where $\tilde{\rho}''_i$ denotes the transposed matrix of $\rho''_i$
(that is, its complex conjugate). This is a legal density matrix for
Bob's subsystem, so that $\sigma$ is a valid density matrix for the
composite system. Consider now an arbitrary $\rho$ whose matrix elements
$\rho_{m\mu,n\nu}$ are given (Latin indices refer to Alice's basis,
Greek indices to Bob's). We can likewise define a partly transposed
matrix

\beq \sigma_{m\mu,n\nu}=\rho_{m\nu,n\mu}. \label{sigma}\eeq
We have just seen that if $\rho$ is separable as in \Eq{separ}), then
$\sigma$ is a legal density matrix. In particular all its
eigenvalues are non-negative.  Thus, if we happen to find that $\sigma$
has a negative eigenvalue, this means that $\rho$ is not separable.

The following proposition will now be proved: {\it If a state $\rho$
satisfies (violates) a Bell inequality, then the partly transposed
$\sigma$ satisfies (violates) the same Bell inequality.\/}\\[2mm]
{\bf Proof: } The Bell inequality (\ref{PF}), combined with the quantum
mechanical rule for computing probabilities, implies that

\beq \sum_K F^K\;\mbox{\rm Tr}\,(\rho\,A_K\otimes B_K)\geq0, \eeq
where $A_K$ and $B_K$ denote projection operators, corresponding to the
detectors of Alice and Bob that are involved in a coincidence of type
$K$. Explicitly,

\beq \sum_K F^K \sum_{mn\mu\nu}\rho_{m\mu,n\nu}\,
   (A_K)_{nm}\,(B_K)_{\nu\mu} \geq0,\label{Bellineq} \eeq
and therefore, by \Eq{sigma}),

\beq \sum_K F^K \sum_{mn\mu\nu}\sigma_{m\nu,n\mu}\,
   (A_K)_{nm}\,(B_K)_{\nu\mu} \geq0. \eeq
Exchanging the indices $\mu$ and $\nu$, we obtain

\beq \sum_K F^K \sum_{mn\mu\nu}\sigma_{m\mu,n\nu}\,
   (A_K)_{nm}\,(B_K)_{\mu\nu} \geq0. \eeq
Comparison with \Eq{Bellineq}) shows that the projection operator $B_K$
has been replaced by its transpose $\tilde{B}^{\phantom{*}}_K\equiv
B^*_K$, which is also a projection operator, corresponding to another
type of detector (the one sensitive to the time-reversed state of Bob's
subsystem~[23]). This concludes the proof of the above proposition. Note
that the latter is valid even if $\sigma$ has a negative eigenvalue and
is not a legal density matrix. We thus see how tenuous the relationship
is between quantum separability and Bell's inequalities.

Similar properties hold if there are more than two observers. Partial
transpositions are defined with respect to any one of them, or to any
subset of them. A necessary (but not a sufficient) condition for quantum
separability is that all these partly transposed density matrices have
non-negative eigen\-values.

The weakness of Bell's inequalities as a condition for quantum
separability is due to the fact that the only use made of the density
matrix $\rho$ is for computing the probabilities of the outcomes of
tests performed on subsystems of a composite system, following a
specific preparation of the latter. An experimental verification of
these inequalities necessitates observing many such composite systems,
all prepared in the same way.  However, if many systems are actually
available, we may also test them collectively, for example two by two,
or three by three, etc., rather than one by one. If we do that, we are
using, instead of $\rho$ (the density matrix of a single system), a new
density matrix, which is $\rho\otimes\rho$, or
\mbox{$\rho\otimes\rho\otimes\rho$,} in a higher dimensional space. We
may then find that there are some density matrices $\rho$ that satisfy
all Bell's inequalities when each system is tested individually, but for
which $\rho\otimes\rho$, or $\rho\otimes\rho\otimes\rho$, etc., violate
some new Bell inequality, because the collective tests require new
experimental setups, and the number of dimensions of Hilbert space
increases~[24]. This property leads to the notion of quantum
distillation~[25--27].

The essence of quantum distillation lies in the reduction of the
effective number of dimensions of Hilbert space by the selection of a
suitable subensemble out of the original quantum ensemble. The
resulting composite systems have lower dimension (e.g., they may be
pairs of spin-$1\over2$ particles in a singlet state). It is then easier
to display a violation of Bell's inequality. The distillation process
involves only local operations by the distant observers and classical
communication between them, but no transfer of {\it quantum\/} systems
(recall that Bell inequalities are fundamentally classical). An example
was given in Sect.~3.3: there  were three observers, one of them
performed a test, and the two other observers retained their subsystems
for further use if, and only if, that test was successful. An even
simpler example was constructed by Popescu~[28] who showed how to
exhibit the hidden nonlocality of a pair of spin-2 systems by selecting
a subensemble whose Hilbert space had effectively two dimensions,
instead of five. More generally, distillation involves taking several
entangled pairs (or trios, etc.), letting the particles held by each
observer interact in an appropriate way, testing a subset of them at
each location, and retaining only the samples that passed all these
tests.

Note that in the above scenario the observers have to follow a definite
protocol. They have no free choice between different experimental
setups. It is only after the distillation process has been brought to
completion that the observers (or a subset of them) have to consider
alternative setups in order to formulate Bell inequalities. The
distillation protocol with two observers follows a sequential method
which has many similarities with the procedure that was specified in
Sect.~3.3 for three or more observers. We can in fact replace Alice and
Bob by several distinct observers, one for each stage of the
distillation (let us call them Alice~I, Alice~II, etc., and Bob~I,
Bob~II, \ldots ). We can then imagine multi\-dimensional diagrams
generalizing Fig.~9, with one axis for each member of the dynasties of
Alice and Bob. The assumption of locality (among different Alices) is
replaced by the assumption that the result of a test performed by
Alice~I does not depend on the choice of the tests that will eventually
be performed by the following Alices (in particular, by the last one,
who will examine the validity of a Bell inequality). It is only the last
step that involves counter\-factual definiteness; there is nothing
counter\-factual in all the preliminary ones.

The reduction of the dimensionality of Hilbert space is due to the
factorization of the states obtained at each distillation stage. For
example, $\psi_{123}$ in \Eq{qm}) becomes

\beq \psi'_{123}=u_1\,\otimes\psi'_{23}, \eeq
where $\psi'_{23}$ is given by \Eq{psi23}). From this point on, we can
safely ignore particle~1, and consider only~2 and~3; hence the reduction
of the number of dimensions.

Note that there exist inseparable quantum states that cannot be
distilled into singlets. In particular, quantum states $\rho$ whose
partial transpose $\sigma$ has no negative eigenvalue have that
property~[29]. Thus, if the preceding conjectures are correct, it
follows that these peculiar inseparable quantum states violate no Bell
inequality, and therefore, owing to Farkas's lemma, their statistical
properties are compatible with the existence of local objective
variables.\bigskip

\noindent{\bf5. \ CRITERIA FOR LOCAL OBJECTIVE VARIABLES}\medskip

\noindent The purpose of this work was to produce a complete set of Bell
inequalities, from which any other Bell inequality can be obtained as a
convex combination. A graphical method was devised, giving a large
number of Bell inequalities (of the Clauser-Horne type). The question is
whether that set of inequalities is complete. During the course of
this investigation, some conjectures
naturally arose. They are illustrated by Figs.~6, 8, and~9.

Expressed in plain words, Fig.~6 says that if an observer has several
detectors, not just one detector as required for formulating the CH
inequality, then that observer has to consider each possible subset of
these detectors as if it were a single detector, for the purpose of
computing the probabilities that appear in the CH inequality. This
conjecture appears quite plausible. Finding a formal proof is proposed
as a challenge to mathematical physicists who are better experts than me
in convex analysis.

Another conjecture, illustrated by Fig.~8, was that if an observer has
more than two alternative setups, it is sufficient to consider them
pairwise in order to write Bell inequalities.  Although this claim
appeared to be plausible, there are counter\-examples~[30], and a simple
algorithm still has to be found for the construction of Farkas vectors
involving more than four sectors in a nontrivial way.

Still another plausible conjecture (Fig.~9) is that if there are $N>2$
observers, only two of them have to consider alternative setups. The
$(N-2)$ other ones perform a single test, which decides whether all the
observers proceed with their parts of the composite system, or they
discard these parts and turn to examine another composite system.
However, I have no proof of this conjecture: it may well be that
there are inseparable quantum states that cannot be distilled, and yet
these states violate a Bell inequality which requires combining data
from more than two observers.

The above discussion suggests that the partial
transposition criterion~[22] is a necessary and sufficient condition for
the compatibility of local objective variables with experimental
results, irrespective of whether or not the composite system is
separable from the point of view of quantum theory. This would be an
interesting link between classical and quantum physics.

It thus seems that there are three different levels of nonlocality in
physics. There are separable quantum systems whose statistical
properties can be mimicked by local objective variables; there are
inseparable quantum systems, which are incompatible with local hidden
variables (either when they are tested singly, or collectively, or when
they are distilled); and there also is an intermediate level:
inseparable systems with positive partial transposition, which cannot be
prepared by local quantum operations and classical communication, but
whose statistical properties, once they have been prepared, do not
conflict with local objective variables.

\bigskip\noindent{\bf ACKNOWLEDGMENTS}\medskip

\noindent It is a pleasure to dedicate this article to Danny
Greenberger, on the occasion of his 65th birthday. I am grateful to
Louis Michel for patiently educating me on the geometry of convex
polytopes, and to Tal Mor for suggesting that the Horodecki states~[29]
might obey every Bell inequality. I also received helpful advice from
Micha\l, Pawe\l, and Ryszard Horodecki, from David Mermin, and from
Abner Shimony. This work was supported by the Gerard Swope Fund and the
Fund for Encouragement of Research.\bigskip  

\noindent{\bf APPENDIX: ALGEBRAIC CONSTRUCTION OF FARKAS
VECTORS}\medskip

\noindent Recall that the vectors $\bB^\1$ (with components $B^\1_K$)
are the extreme points of a convex cone. They span a linear space of
dimension $N_D$ (that is, only $N_D$ of these vectors are linearly
independent). Note that $N_D<N_K$, where $N_K$, given by \Eq{NK}), is
the number of dimensions of the basis used to represent these vectors
(each basis element corresponds to a detection coincidence). The
complementary subspace, of dimension $N_K-N_D$, is spanned by the null
vectors $Z_K$ that were defined in \Eq{ZK}). (In this Appendix, I am
using a Cartesian scalar product, and there is no reason to distinguish
upper and lower $K$ indices.) 

For example, if there are two observers, and each one has two
alternative experiments, with $A_i$ and $B_j$ distinct outcomes,
respectively, there are

\beq N_Z=\sum_{i=1}^2 A_i+\sum_{j=1}^2 B_j-1\eeq
linearly independent null vectors, that will be denoted as $\bZ^\sigma$,
with $\sigma=1,\ldots,N_Z$. It follows that

\beq N_D=N_K-N_Z=\Bigl(\sum_iA_i-1\Bigr)\,\Bigl(\sum_jB_j-1\Bigr).\eeq
More generally, if Alice and Bob have $S_A$ and $S_B$ alternative
setups, then

\beq N_Z=(S_B-1)\sum_iA_i+(S_A-1)\sum_jB_j-{S_A\choose2}{S_B\choose2}.
\eeq
Similar formulas can be found for the case of more than two observers.

Obviously, the vector

\beq \bC=\sum_\1\bB^\1 \eeq
is in the interior of the cone whose edges are the vectors $\bB^\1$. The
question is whether the vector \bP, whose components are the detection
probabilities $P_K$, also is in the interior. Consider a hypersurface
element formed by a subset of $N_D-1$ different $\bB^\1$.  The vectors
\bC\ and \bP\ are on the same side of that hyper\-surface if the
oriented volumes formed by them and by the subset $\{\bB^\1\}$ have the
same sign. In order to have a simple algebraic criterion for this, it is
convenient to add $N_Z$ dimensions. Consider therefore the
hyper\-surface element of dimension $N_K-1$ formed by the $N_D-1$
vectors of $\{\bB^\1\}$ and the $N_Z$ vectors $\bZ^\sigma$ (recall that
these are two orthogonal subspaces). We can then write two determinants
of order $N_K$ by taking these $N_K-1$ vectors together with \bP\ or
with \bC. Namely, the $K$-th row or column of these determinants is

\beq P_K\cdots B^\1_K\cdots Z^\sigma_K\cdots\qquad\qquad\mbox{\rm or}
\qquad\qquad C_K\cdots B^\1_K\cdots Z^\sigma_K\cdots.\eeq
The question is whether these determinants have the same sign or, in
other words, whether their product is positive. Since the product of two
determinants is equal to the determinant of the product of their
matrices, let us multiply the matrices. Let $K$ be considered as a
column index in the first term in (\theequation), and as a row index in
the second one, so that the matrix product involves a sum over the index
$K$. Recall that

\beq \bZ^\sigma\cdot\bB^\1=\bZ^\sigma\cdot\bC=\bZ^\sigma\cdot\bP=0,\eeq
where the last equality is \Eq{PF0}) in the present notations. Then the
product of the two above determinants factorizes as

\beq \det\left|\begin{array}{cccc}
 \bP\cdot\bC & \cdots & \bP\cdot\bB^\mu & \cdots \\
 \vdots & \vdots & \vdots & \vdots \\ 
 \bB^\1\cdot\bC & \cdots & \bB^\1\cdot\bB^\mu & \cdots \\
 \vdots & \vdots & \vdots & \vdots \end{array}
 \right| \quad\times\quad\det\,\Bigl|\bZ^\sigma\cdot\bZ^\tau\Bigr|.
\label{dets} \eeq
In the first determinant in (\theequation), the indices $\1$ and $\mu$
denote rows and columns, respectively. The second determinant in
(\theequation) is always positive (this is a generalization of the
Schwarz inequality, easily proved by noting that the matrix of that
determinant is a sum of positive matrices). Therefore the requirement
that \bC\ and \bP\ lie on the same side of the hyperplane formed by
$\{\bB^\1\}$ implies that the first determinant in (\theequation) is
positive. Expanding that determinant from its first row, we obtain an
inequality which can be written as

\beq \sum_K P_K\,G^K>0. \label{PG} \eeq
Note that all the terms appearing in (\ref{dets}) are integers,
therefore the coefficients of $P_K$ in Eq.~(\theequation) are also
integers.

However, the $G^K$ defined above are not, in general, the components
of a Farkas vector if the subset $\{\bB^\1\}$ is chosen arbitrarily.
This subset is relevant only if it forms a {\it face\/} of the convex
cone. Namely the subset $\{\bB^\1\}$ must be chosen in such a way that
any $\bB^\kappa$ that does not belong to it lies on the same side as
$\bC$. This condition can be verified by replacing \bP\ in the first
determinant in (\ref{dets}) by each one of the $\bB^\kappa$:

\beq \forall\kappa\qquad\det\left|\begin{array}{cccc}
 \bB^\kappa\cdot\bC & \cdots & \bB^\kappa\cdot\bB^\mu & \cdots\\
 \vdots & \vdots & \vdots & \vdots \\ 
 \bB^\1\cdot\bC & \cdots & \bB^\1\cdot\bB^\mu & \cdots\\
 \vdots & \vdots & \vdots & \vdots
 \end{array}\right|>0. \label{face} \eeq
If this inequality holds for all the $\bB^\kappa$ that do not belong to
$\{\bB^\1\}$ (which is the same subset as $\{\bB^\mu\}$, of course),
then this subset defines a face of the convex cone (subsets that do not
satisfy this condition are hyperplanes interior to the cone). It is only
for these faces that the vectors $G^K$, as defined above, are Farkas
vectors. The set of Farkas vectors obtained in this way is manifestly
complete. It may however be overcomplete, because some of these vectors
may not be extremal.

Note that 

\beq \bB^\1\cdot\bC=\sum_\nu\bB^\1\cdot\bB^\nu.\eeq 
It is enough to include in that sum the $\nu$ that do not belong to the
set of $\mu$ that appears in \Eq{face}), because we can always subtract
from the first column all the other ones, without changing the value of
the determinant.

Finally, we note that the symmetric tensor

\beq A^{\1\mu}=\bB^\1\cdot\bB^\mu=\sum_K B^\1_K\,B^\mu_K \eeq
appears frequently in these calculations. Owing to \Eq{Bfactor}),
$A^{\1\mu}$ can be factorized. For example, if $\1=[bsu;\beta\rho]$ as
in \Eq{Bfactor}), and if $\mu=[crv;\alpha\tau]$, we obtain the simple
result

\beq  A^{\1\mu}=(\delta_{bc}+\delta_{rs}+\delta_{uv})\,
 (\delta_{\alpha\beta}+\delta_{\rho\tau}). \eeq
In particular, $A^{\1\1}=6$, which is the number of sectors in the
diagram corresponding to \Eq{Bfactor}).

It is not difficult to write a computer program that tests the signs of
all these determinants. Their absolute values are not needed until we
have found a face. Only then we have to compute the subdeterminants that
give the components $G^K$ for \Eq{PG}). The only difficulty here is that
the number of subsets $\{\bB^\1\}$ that have to be examined, to decide
whether they are faces of the convex cone, increases exponentially as
$N_\1\choose N_D-1$. The problem is known to be NP-complete~[31]. It is
likely that progress can be done by using the symmetries of the set of
$B^\1_K$. This will be a subject for future research.\bigskip

\noindent{\bf REFERENCES}

\frenchspacing \begin{enumerate}
\item J. S. Bell, Physics (N. Y.) {\bf 1}, 195 (1964).
\item J. F. Clauser, M. A. Horne, A. Shimony, and R. A. Holt, Phys. Rev.
Letters {\bf 23}, 880 (1969).
\item A. Peres, {\it Quantum Theory: Concepts and Methods\/}
(Kluwer, Dordrecht, 1993) Chap. 6 and references therein.
\item J. F. Clauser and M. A. Horne, Phys. Rev. D {\bf 10}, 526 (1974).
\item N. D. Mermin, Am. J. Phys. {\bf 49}, 940 (1981).
\item J. S. Bell, in {\it Foundations of Quantum Mechanics, Proceedings
of the International School of Physics ``Enrico Fermi,'' Course XLIX\/},
ed. by B.~d'Espagnat (Academic, New York, 1971).
\item E. Schr\"odinger, Naturwiss. {\bf23}, 800, 823, 844 (1935)
[transl. in {\it Quantum Theory and Measurement\/} ed. by J. A. Wheeler
and W. H. Zurek (Princeton University Press, Princeton, 1983) p. 152].
\item A. Garg and N. D. Mermin, Found. Phys. {\bf 14}, 1 (1984).
\item I. Pitowsky, Brit. J. Phil. Sci. {\bf 45}, 95 (1994).
\item R. T. Rockafellar, {\it Convex Analysis\/} (Princeton Univ. Press,
Princeton, 1970), p.~200.
\item S. L. Braunstein and C. M. Caves, Ann. Phys. (NY) {\bf202}, 22
(1990).
\item A. Garg and N. D. Mermin, Phys. Rev. Letters {\bf49}, 1220 (1982).
\item N. D. Mermin, Philos. Sci. {\bf50}, 359 (1983).
\item P. Horodecki and R. Horodecki, Phys. Rev. Letters {\bf76}, 2196
(1996).
\item N. D. Mermin and A. Garg, Phys. Rev. Letters {\bf76}, 2197 (1996).
\item R. Horodecki, P. Horodecki, and M. Horodecki, Physics Letters A
{\bf200}, 340 (1995).
\item N. D. Mermin, Phys. Rev. Letters {\bf 65}, 1838 (1990).
\item M. Ardehali, Phys. Rev. A {\bf46}, 5375 (1992).
\item N. Gisin and H. Bechmann-Pasquinucci, Physics Letters A {\bf246},
 1 (1998).
\item S. L. Braunstein and A. Mann, Phys. Rev. A {\bf47}, 2427 (1993).
\item R. F. Werner, Phys. Rev. A {\bf40}, 4277 (1989).
\item A. Peres, Phys. Rev. Letters {\bf77}, 1413 (1996).
\item P. Busch and P. Lahti, Found. Phys. Letters {\bf10}, 113 (1997).
\item A. Peres, Phys. Rev. A {\bf54}, 2685 (1996).
\item C. H. Bennett, H. J. Bernstein, S. Popescu, and B. Schumacher,
Phys. Rev. A {\bf53}, 2046 (1996).
\item C. H. Bennett, G. Brassard, S. Popescu, B. Schumacher, J. A.
Smolin, and W.  K. Wootters, Phys. Rev. Letters {\bf76}, 722 (1996).
\item D. Deutsch, A. Ekert, R. Jozsa, C. Macchiavello, S. Popescu, and
A. Sanpera, Phys. Rev. Letters {\bf77}, 2818 (1996).
\item S. Popescu, Phys. Rev. Letters {\bf74}, 2619 (1995).
\item M. Horodecki, P. Horodecki, and R. Horodecki, Phys. Rev. Letters
{\bf80}, 5239 (1998).
\item A. Garg and N. D. Mermin, Phys. Rev. D {\bf27}, 339 (1983),
Sect.~III.
\item I. Pitowski, Math. Programming {\bf50}, 395 (1991).
\end{enumerate}\nonfrenchspacing\clearpage

\noindent FIG. 1. \ Alice has a choice of three tests, whose outcomes
label the vertical axis. Bob has a choice of two tests (horizontal
axis). Each one of the six sectors thus represents a possible joint
experiment. A definite value of the hidden variable $\1$ determines the
outcomes of all these experiments, as shown by the pattern of dotted
lines. Each node in that pattern represents a coincidence, labelled by
an index $K$. \bigskip

\noindent FIG. 2. \ Same as Fig.~1, with three observers. The axes are
labelled by the various results that can be obtained by each one of
them. Each value of $\1$ determines on each face a pattern similar to
the one in Fig.~1. The intersections of all the planes formed by these
patterns are the detection coincidences that correspond to that value of
$\1$.\bigskip

\noindent FIG. 3. \ Two examples of null Farkas vectors $Z^K$. A black
cell means $-1$, a gray cell means +1. Empty cells are zeros.\bigskip

\noindent FIG. 4. \ Farkas vector for the Clauser-Horne inequality. A
black cell means $-1$, a gray cell means +1. Empty cells are zeros.
\bigskip

\noindent FIG. 5. \ This Farkas vector can be reduced to one without
negative elements by adding null vectors as those of Fig.~3.\bigskip

\noindent FIG. 6. \ Alice has a choice of two tests, one with 5 answers
and one with 4 (each row of the diagram corresponds to one answer). Bob
has a choice of two tests, with 6 and 5 answers, respectively (one
column for each answer). There are 781\,200 different Farkas vectors
(each one with 99 components) which can be obtained as in Fig.~4 by
lumping together the outputs of some detectors.\bigskip

\noindent FIG. 7. \ Four equivalent Farkas vectors. They differ by the
addition of null vectors, like those of Fig.~3.\bigskip

\noindent FIG. 8. \ The Farkas vector for a chained CH inequality can be
decomposed into a sum of two Farkas vectors for ordinary CH
inequalities. Recall that a black cell means $-1$, a gray cell means +1,
and empty cells are zeros.\bigskip

\noindent FIG. 9. \ Farkas vector for three observers: there are 64
components, represented by an array of cubes. The black cube stands for
$-1$, the three gray cubes for +1, and the 60 other components of the
Farkas vector vanish. Note that all the nonvanishing components belong
to the same slab of the cubic array.
\end{document}